\documentclass{aa}
\newcommand{\kA}{k^{\rm A}}
\newcommand{\kc}{k^{\rm c}}
\usepackage{graphicx}
\usepackage{psfrag}
\usepackage{url}
\newcommand{\EQ}{\begin{equation}}
\newcommand{\EN}{\end{equation}}
\newcommand{\EQA}{\begin{eqnarray}}
\newcommand{\ENA}{\end{eqnarray}}
\newcommand{\eq}[1]{(\ref{#1})}
\newcommand{\EEq}[1]{Equation~(\ref{#1})}
\newcommand{\Eq}[1]{Eq.~(\ref{#1})}
\newcommand{\Eqs}[2]{Eqs~(\ref{#1}) and~(\ref{#2})}

\newcommand{\Eqss}[2]{Eqs~(\ref{#1})--(\ref{#2})}

\newcommand{\Fig}[1]{Fig.~\ref{#1}}
\newcommand{\FFig}[1]{Figure~\ref{#1}}

\newcommand{\bra}[1]{\langle #1\rangle}

\newcommand{\meanB}{\overline{B}}

\newcommand{\meanBB}{\overline{\vec{B}}}

\newcommand{\meanUU}{\overline{\vec{U}}}

\newcommand{\meanqq}{\overline{\vec{q}}}

\newcommand{\meanuu}{\overline{\mbox{\boldmath $u$}}{}}{}
{}
{}
{}
{}
{}
\newcommand{\zz}{\hat{\mbox{\boldmath $z$}} {}}

\newcommand{\xx}{{\vec{x}}}

\newcommand{\qq}{{\vec{q}}}

\newcommand{\UU}{{\vec{U}}}
\newcommand{\uu}{{\vec{u}}}
\newcommand{\BB}{{\vec{B}}}
\newcommand{\JJ}{{\vec{J}}}

\newcommand{\aaaa}{{\vec{a}}}
\newcommand{\bb}{{\vec{b}}}

\newcommand{\ff}{\mbox{\boldmath $f$} {}}

\newcommand{\FF}{{\vec{F}}}

\newcommand{\kk}{{\vec{k}}}

\newcommand{\grav}{\mbox{\boldmath $g$} {}}
\newcommand{\nab}{\vec{\nabla}}
\newcommand{\OO}{\mbox{\boldmath $\Omega$} {}}

\newcommand{\LLLL}{\mbox{\boldmath ${\sf L}$} {}}

\newcommand{\ii}{{\rm i}}

\newcommand{\DD}{{\rm D} {}}

\newcommand{\dd}{{\rm d} {}}
\newcommand{\const}{{\rm const}  {}}

\newcommand{\ea}{{\rm et al.\ }}

\def\half{{\textstyle{1\over2}}}

\def\quarter{{\textstyle{1\over4}}}
\newcommand{\yapj}[3]{ #1, {ApJ,} {#2}, #3}

\newcommand{\yana}[3]{ #1, {A\&A,} {#2}, #3}

\newcommand{\yanar}[3]{ #1, {A\&AR,} {#2}, #3}

\newcommand{\yprl}[3]{ #1, {PRL,} {#2}, #3}
\newcommand{\ypre}[3]{ #1, {PRE,} {#2}, #3}

\newcommand{\ymn}[3]{ #1, {MNRAS,} {#2}, #3}

\newcommand{\yjour}[4]{ #1, {#2}, {#3}, #4}

\newcommand{\ybook}[3]{ #1, {#2} (#3)}

\newcommand{\sana}[1]{ #1, {A\&A} (submitted)}

\begin{document}

\input{epsf}
\title{Nonaxisymmetric stability in the shearing sheet approximation}
\author{A.\ Brandenburg\inst{1} \and B.\ Dintrans\inst{2}}
\institute{NORDITA, Blegdamsvej 17, DK-2100 Copenhagen \O, Denmark \and
Observatoire Midi-Pyr\'en\'ees, CNRS et Universit\'e Toulouse 3,
14 avenue Edouard Belin, 31400 Toulouse, France}

\offprints{brandenb@nordita.dk}

\date{\today,~ $ $Revision: 1.82 $ $}

\abstract{}
{To quantify the transient growth of nonaxisymmetric perturbations
in unstratified
magnetized and stratified non-magnetized rotating linear shear flows
in the shearing sheet approximation of accretion disc flows.}
{The Rayleigh quotient in
modal approaches for the linearized equations (with time-dependent
wavenumber) and the amplitudes from
direct shearing sheet simulations using a finite
difference code are compared.}
{Both approaches agree in their predicted growth behavior.
The magneto-rotational instability for axisymmetric and non-axisymmetric
perturbations is shown to have the same dependence of the (instantaneous)
growth rate on the wavenumber along the magnetic field,
but in the nonaxisymmetric case the growth is only transient.
However, a meaningful dependence of the Rayleigh quotient on the
radial wavenumber is obtained.
While in the magnetized case the total amplification factor can be
several orders of magnitude, it is only of order ten or less in the
nonmagnetic case.
Stratification is shown to have a stabilizing effect.
In the present case of shearing-periodic boundaries the (local) 
strato-rotational
instability seems to be absent.}
{}

\maketitle

\section{Introduction}

The gas in accretion discs is generally though to be turbulent
(Shakura \& Sunyaev 1973).
This allows potential energy to be converted into kinetic energy
which can then be dissipated (e.g., Frank et al.\ 1992).
This in turn can lead to significant amounts of observable radiation.
To sustain the turbulence, there has to be some instability.
This instability is now generally thought to be the Balbus-Hawley
or magneto-rotational instability (hereafter referred to as MRI;
see Balbus \& Hawley 1998 for a review).
This instability is linear and local in that it does not rely on the
presence of boundaries.
It exists already in the axisymmetric case in the presence of an
external vertical field, which makes this instability technically
easy to study.

In the nonaxisymmetric case, an azimuthal pattern will be sheared out
differentially, i.e.\ patterns in the inner parts of the disc are advected
faster than in the outer parts.
After an increasing number of orbits, this causes more and more rapid
variations in the radial direction.
Loosely speaking, this makes the radial wavenumber time-dependent.

In the presence of radial boundaries or other radial non-uniformities,
and certainly also in the presence of nonlinearity,
any radial pattern that was initially harmonic will become anharmonic.
This produces a spectrum of wavenumbers even for monochromatic initial
conditions.
Mathematically, the stability of such a problem can be studied by
solving a one-dimensional eigenvalue problem subject to radial boundary
conditions (Ogilvie \& Pringle 1996).
Even though boundary conditions are important in that case, the results may still
be relevant for driving turbulence provided the instability is not
limited to the vicinity of the boundaries.

In many numerical simulations of accretion disc turbulence the shearing
sheet (or shearing box) approximation has been employed
(Hawley et al.\ 1995, Matsumoto \& Tajima 1995, Brandenburg et al.\ 1995).
This approximation represents the other extreme, where boundaries and
any non-uniformities are strictly removed.
Any instability that survives under these conditions is often referred
to as a ``local'' instability, even though its onset properties
may depend on the system size (as is typical of all long-wavelength
instabilities).

For nonaxisymmetric solutions, a purely analytic treatment of the shearing
sheet model is generally impossible, because the solutions exhibit a
complicated temporal behavior that cannot even be approximated by an
exponential time evolution.
In fact, all nonaxisymmetric solutions only exhibit transient growth
whose speed of growth depends on the instantaneous wavevector.
However, for the MRI it turns out that the Rayleigh quotient obtained
from the time-dependent nonaxisymmetric solution is a good approximation
to the usual eigenvalue in the much simpler axisymmetric problem.
The purpose of the present paper is to attempt a more systematic
survey of the nonaxisymmetric problem by studying the dependence of
Rayleigh quotient in different two-dimensional parameter planes.
We begin with the fairly well understood MRI and then turn to
the less well understood problem with vertical density stratification,
but no magnetic field.
The latter case is less well understood, although it has been shown
that in the presence of radial boundaries there is a linear instability
of the Taylor--Couette problem with density stratification along the
axis (Molemaker et al.\ 2001, Shalybkov \& R\"udiger 2005, Umurhan 2005).
This ``strato-rotational'' instability (hereafter refereed to as SRI)
has been confirmed numerically in the presence of boundaries
(Brandenburg \& R\"udiger 2006), but it
may still be local in character, i.e.\ its properties may not be
sensitive to the presence of boundaries (Dubrulle et al.\ 2005).

It turns out that in the nonaxisymmetric case without boundaries
there can at most only be transient growth.
This is also true of the MRI, which has an instantaneous growth rate
quite analogous to that in the axisymmetric case.
In shearing sheet simulations, sustained instability can only be
the result of nonlinearity allowing mode coupling and hence the
recycling of energy into new modes viable of repeated growth.

\section{The shearing sheet formalism}

The full hydrodynamic and
magnetohydrodynamic equations can always be written in the form
\EQ
{\DD\qq\over\DD t}=\FF(\qq),
\label{DQDt}
\EN
where $\qq$ is a state vector combining all components of velocity,
density, entropy, and the magnetic field, and
$\DD/\DD t=\partial/\partial t+\UU\cdot\nab$ is the advective derivative.
The velocity $\UU$ can be subdivided into an equilibrium solution
(the mean flow $\meanUU$) and the departures from the mean flow, so
\EQ
\UU=\meanUU+\uu.
\label{UUsplit}
\EN
In the shearing sheet approximation, the mean flow depends linearly
on the cross-stream coordinate, say $x$, so we assume the mean flow
to be
\EQ
\meanUU=(0,Sx,0)^T,
\label{MeanFlow}
\EN
where $S$ denotes the gradient of the shear flow.
In a local model of a keplerian disc we have $S=-{3\over2}\Omega$,
where $\Omega$ is the local angular velocity.

Inserting \Eq{MeanFlow} in \Eq{DQDt} yields
\EQ
{\partial\qq\over\partial t}+Sx{\partial\qq\over\partial y}
+\uu\cdot\nab\qq=\FF(\qq),
\label{DQDt2}
\EN
where the second term on the left hand side has an explicit $x$
dependence.
It turns out that none of the other terms have an explicit $x$ dependence.
This is because we are restricting ourselves only to {\it linear}
shear flows, so the flow has constant gradients.

\EEq{DQDt2} can be linearized with respect to the departures
from the equilibrium solution, $\meanqq$, so we write
$\qq=\meanqq+\qq'$ and have
\EQ
{\partial\qq'\over\partial t}+Sx{\partial\qq'\over\partial y}=\LLLL\qq',
\label{dqdt}
\EN
where $\LLLL$ is a matrix with differential operators and constant
coefficients.
\EEq{dqdt} can be solved by making the ``shearing sheet'' ansatz
(Goldreich \& Lynden-Bell 1965, Balbus \& Hawley 1992a)
\EQ
\qq'(x,y,z,t)=\hat{\qq}(t)\,
\exp\left[\ii k_x(t)x+\ii k_yy+\ii k_zz\right].
\label{expansion}
\EN
Note that by differentiating \Eq{expansion} with respect to $t$, one pulls
down a term proportional to $\ii(\dd{k_x}/\dd t)x$.
This explicitly $x$-dependent term can be arranged to cancel the
second term of \Eq{dqdt} by choosing
\EQ
k_x(t)=k_{x0}-k_ySt.
\label{kxt}
\EN
This leads to a set of {\it ordinary} differential equations,
\EQ
{\dd\hat{\qq}\over\dd t}=\hat{\LLLL}\hat{\qq},
\label{ode}
\EN
where $\hat{\LLLL}$ is a matrix with coefficients that are independent
of $\xx$ and depend at most only on $t$.

In practice, we solve \Eq{ode} numerically and monitor the evolution
of the norm, $\bra{\hat{\qq}|\hat{\qq}}$, and of the Rayleigh quotient
\EQ
\lambda(t)={\bra{\hat{\qq}|\hat{\LLLL}\hat{\qq}}\over\bra{\hat{\qq}|\hat{\qq}}},
\label{deflambdat}
\EN
where $\displaystyle \bra{\aaaa|\bb}=\sum_{i=1}^Na^*_ib_i$ defines
a scalar product, and $N$ is the rank of the matrix $\hat{\LLLL}$.
We recall that, if the matrix was independent of $t$, the
Rayleigh quotient would be between the largest and the smallest
eigenvalue of $\hat{\LLLL}$.

In the following we discuss first the MRI and turn then to the
case with stratification and address the possibility of the
strato-rotational instability (SRI).

\section{MRI}

\subsection{Basic equations}

Stratification is unimportant for the MRI, so we focus on the simple
case with uniform background density.
To simplify the problem further, we assume an isothermal equation of
state, so the pressure is given by $p=c_{\rm s}^2\rho$.
Here, $c_{\rm s}$ is the sound speed which is assumed constant.
The full set of equations, in the presence of rotation with
angular velocity $\OO$, is then
\EQ
\rho{\DD\UU\over\DD t}=-c_{\rm s}^2\nab\rho-2\OO\times\rho\UU
-\rho\nab\psi+\JJ\times\BB,
\label{dUdt}
\EN
\EQ
{\DD\BB\over\DD t}=\BB\cdot\nab\UU-\BB\nab\cdot\UU,
\label{dBdt}
\EN
\EQ
{\DD\rho\over\DD t}=-\rho\nab\cdot\UU,
\label{drhodt}
\EN
where $\DD/\DD t=\partial/\partial t+\UU\cdot\nab$ is the
advective derivative with respect to the total flow velocity,
$\BB$ is the magnetic field,
$\JJ=\nab\times\BB/\mu_0$ is the current density,
and $\mu_0$ is the vacuum permeability.

In an accretion disc, $\psi=\half\Omega^2(3x^2-z^2)$ is the tidal
potential that is derived by linearizing the gravitational potential
with respect to some point in the midplane of the disc some distance
away from the central object.
However, in the following we ignore vertical gravity and assume a
more general body force giving rise to the shear flow, so we assume
$\psi=\Omega Sx^2$.
The equilibrium solution is then given by $\meanUU=(0,Sx,0)$,
$\meanBB=\const$, and $\rho=\const$.
The mean flow is obtained by balancing
$2\OO\times\meanUU$ against $\nab\psi$.
Inserting \Eq{UUsplit} into the $\UU\cdot\nab\UU$ nonlinearity of
the momentum equation, we obtain
\EQ
\UU\cdot\nab\UU=\meanUU\cdot\nab\uu+\uu\cdot\nab\meanUU+\uu\cdot\nab\uu,
\EN
where $\meanUU\cdot\nab\meanUU=0$ has been used.
Note also that the term $\uu\cdot\nab\meanUU=(0,Su_x,0)$ can be combined with
the Coriolis force $-2\OO\times\uu$ to give the force
\EQ
\ff(\uu)=\pmatrix{2\Omega u_y\cr-(2\Omega+S) u_x\cr0},
\label{ffuuDefn}
\EN
which describes epicyclic deviations from purely circular motion.
In terms of the departures from the mean flow, $\uu$, the momentum
equation \eq{dUdt} can then be written as (Brandenburg et al.\ 1995)
\EQ
\rho{\DD\uu\over\DD t}=\rho\ff(\uu)-c_{\rm s}^2\nab\rho+\JJ\times\BB.
\label{dudt}
\EN
Likewise, in the induction equation, there is the stretching term
$\BB\cdot\nab\UU$ on the right hand side of the induction equation
\eq{dBdt}, which leads to a term $\BB\cdot\nab\meanUU=(0,SB_x,0)$.
We emphasize again that in the two expressions $(0,Su_x,0)$ and
$(0,Sb_x,0)$, shear only introduces terms with constant coefficients.

So far, all equations have been fully nonlinear.
We can now linearize the equations about $\meanuu=0$,
$\meanBB={\rm const}$ and denote the departures from the
equilibrium solution by a prime.
The linearized Lorentz force can be written in the form
\EQ
(\nab\times\BB')\times\meanBB
=\meanBB\cdot\nab\BB'-\nab(\meanBB\cdot\BB').
\EN
In the following two subsections we consider the cases of imposed
fields that point either in the vertical or in the azimuthal directions.
The former case is particularly instructive, because it allows an
instability already in the much simpler axisymmetric case.

\subsection{Vertical field and axisymmetric perturbations}

\begin{figure}[t!]
\centering\includegraphics[width=\columnwidth]{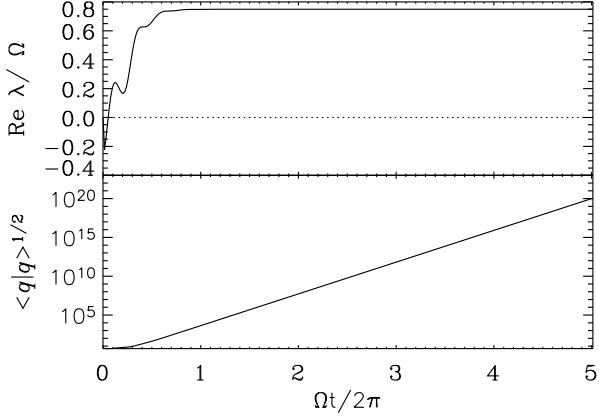}
\caption{Axisymmetric MRI: evolution of the real part of $\lambda$
(upper panel) and the norm $\bra{q|q}^{1/2}$
(or gain; see lower panel), for $v_{\rm A}k_z=1$ using a vertical
initial field pointing in the $z$ direction,
with $k_x=k_y=0$ and $S=-{3\over2}\Omega$.}
\label{FMRI_axi}
\end{figure}

For a vertical field, $\meanBB=(0,0,\meanB_z)$, we have
\EQ
\meanBB\cdot\nab\BB'-\nab(\meanBB\cdot\BB')
=\meanB_z\pmatrix{
\partial_z B'_x-\partial_x B'_z\cr
\partial_z B'_y-\partial_y B'_z\cr
0},
\EN
and the terms on the right hand side of the linearized induction equation are
\EQ
\meanBB\cdot\nab\uu'-\meanBB\nab\cdot\uu'
=\meanB_z\pmatrix{
\partial_z u'_x\cr
\partial_z u'_y\cr
-\partial_x u'_x-\partial_y u'_y}.
\EN
With these preparations we can write down the matrix $\hat{\LLLL}^{(z)}$
for the MRI with an imposed vertical equilibrium field,
$$
\hat{\LLLL}^{(z)}\!\!=\pmatrix{
0&2\Omega&0&\ii\kA_z&0&-\ii\kA_x&-\ii\kc_x\cr
-2\Omega^{\rm S}&0&0&0&\ii\kA_z&-\ii\kA_y&-\kc_y\cr
0&0&0&0&0&0&-\ii\kc_z\cr
\ii\kA_z&0&0&0&0&0&0\cr
0&\ii\kA_z&0&S&0&0&0\cr
-\ii\kA_x&-\ii\kA_y&0&0&0&0&0\cr
-\ii\kc_x&-\ii\kc_y&-\ii\kc_z&0&0&0&0}\!,
$$
where we have used the abbreviations $2\Omega^{\rm S}=2\Omega+S$,
$\kA_i=k_i v_{\rm A}$ (sometimes also $\kk_{\rm A}$),
$\kc_i=k_i c_{\rm s}$ (or $\kk_{\rm c}$), and the state vector
is $\hat{\qq}=(\hat{v}_x,\hat{v}_y,\hat{v}_z,\hat{b}_x,\hat{b}_y,
\hat{b}_z,\hat{\Lambda})^T$, where the hats denote the shearing sheet
expansion analogously to \Eq{expansion}.
Here we have used the rescaled variables
$\Lambda=c_{\rm s}\ln\rho$ and $\bb=\BB/\sqrt{\mu_0\rho_0}$ to make
$\hat{\LLLL}$ hermitian if $S=0$.
To ensure solenoidality of the magnetic field, we calculate $\hat{b}_x$
for the initial perturbation from (Balbus \& Hawley 1992a)
\EQ
\hat{b}_x=-(k_y\hat{b}_y+k_z\hat{b}_z)/k_{x0}
\quad\mbox{(for $k_{x0}\neq0$)}.
\EN
For $k_x=k_y=0$, the matrix $\hat{\LLLL}$ has an unstable eigenvalue
$\lambda=\half S$ at $k_z={15\over16}\Omega$.
In \Fig{FMRI_axi} we present a solution of the ordinary differential
equation \eq{ode} using $\hat{\LLLL}^{(z)}$ for $S=-{3\over2}\Omega$
and $v_{\rm A}k_z=1$.
Note that ${\rm Re}(\lambda/\Omega)$ approaches ${3\over4}\Omega$,
i.e.\ the most unstable eigenvalue.
This suggests that the Rayleigh quotient $\lambda(t)$ is indeed a
good approximation to the most unstable eigenvalue.
In agreement with earlier work (Balbus \& Hawley 1992a,
Kim \& Ostriker 2000), the maximum growth rate agrees with the Oort
$A$-value (Balbus \& Hawley 1992b), which is $-{1\over2}S$, or
${3\over4}\Omega$ for keplerian rotation.

In the special case $k_y=0$ (axisymmetry), $\hat{\LLLL}^{(z)}$
is independent of $t$ and the solution is given by
\EQ
\hat{\qq}=\tilde{\qq}e^{\lambda t}\quad\mbox{(for $k_y=0$)}.
\EN
This leads to the well-known dispersion relation (Balbus \& Hawley 1991)
\EQ
\left[\lambda^4+\lambda^2(2k_{\rm A}^2+\kappa^2)
+k_{\rm A}^2(k_{\rm A}^2-2\Omega S)\right]
(\lambda^2+k_{\rm c}^2)=0,
\EN
where $\kappa^2=4\Omega\Omega^S$ is the square of the epicyclic frequency.
The resulting growth rates are
\EQ
\lambda_\pm^2=-v_{\rm A}^2\kk^2-\half\kappa^2
\pm\sqrt{4v_{\rm A}^2\kk^2\OO^2+\quarter\kappa^4}.
\EN
In the range $0<v_{\rm A}^2\kk^2<2\Omega S$, $\lambda$
can have real values, where $\lambda_+>0$, corresponding to instability.

\subsection{Azimuthal field, nonaxisymmetric perturbations}

Next we turn to the nonaxisymmetric problem.
Of particular interest is 
the case of a purely azimuthal field $\meanBB=(0,\meanB_y,0)$,
which gives rise to the matrix
$$
\hat{\LLLL}^{(y)}\!\!=\pmatrix{
0&2\Omega&0&\ii\kA_y&-\ii\kA_x&0&-\ii\kc_x\cr
-2\Omega^{\rm S}&0&0&0&0&0&-\kc_y\cr
0&0&0&0&-\ii\kA_z&\ii\kA_y&-\ii\kc_z\cr
\ii\kA_y&0&0&0&0&0&0\cr
-\ii\kA_x&0&-\ii\kA_z&S&0&0&0\cr
0&0&\ii\kA_y&0&0&0&0\cr
-\ii\kc_x&-\ii\kc_y&-\ii\kc_z&0&0&0&0}\!.
$$
Since $\hat{\LLLL}^{(y)}$ is time dependent, the solution $\hat{\qq}$
will not have an exponential time dependence and the eigenvalues
of $\hat{\LLLL}^{(y)}$ cannot be interpreted as growth rate.
Therefore we use the Rayleigh quotient $\lambda$ instead.

\begin{figure}[t!]
\centering\includegraphics[width=\columnwidth]{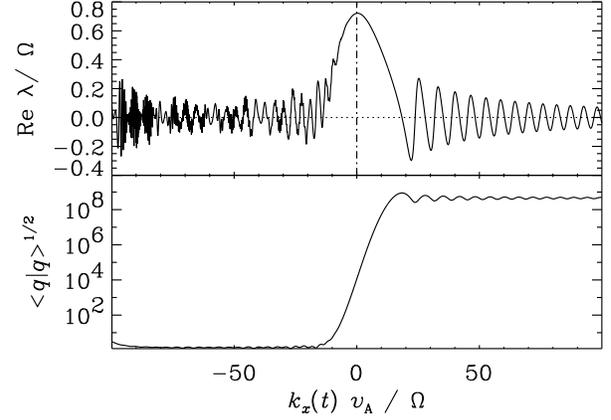}
\caption{Nonaxisymmetric MRI: evolution of the real part of $\lambda$
(upper panel) and the norm $\bra{q|q}^{1/2}$ (lower panel),
for $v_{\rm A}k_y/\Omega=1$  using an azimuthal
initial field pointing in the $y$ direction, and $k_z=0$.}
\label{FMRI_bump}
\end{figure}

\begin{figure}[t!]
\centering\includegraphics[width=\columnwidth]{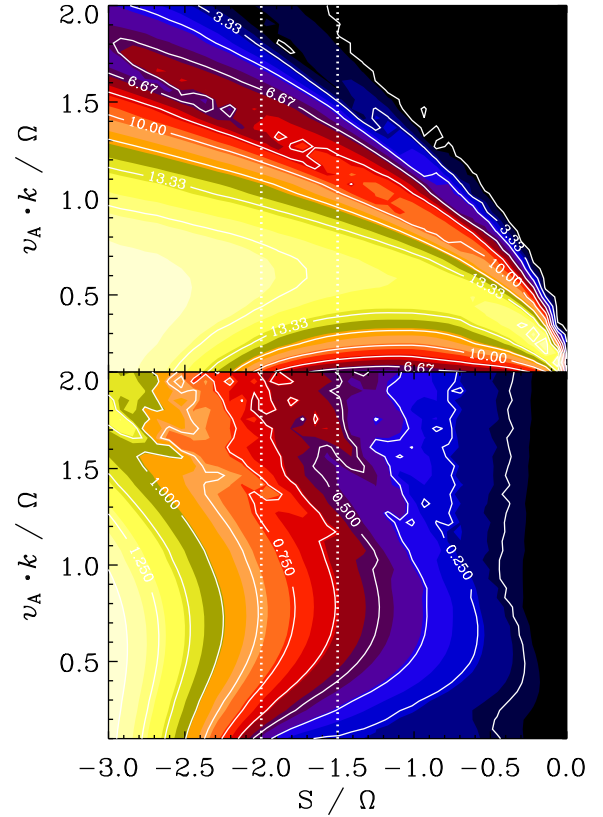}
\caption{Nonaxisymmetric MRI: dependence of the gain factor in
logarithmic representation (upper panel) and of
$\max_t({\rm Re}\lambda)/\Omega$ (lower panel) on $S/\Omega$
and $v_{\rm A}k/\Omega$ for a vertical field.
Note that $\max_t({\rm Re}\lambda)/\Omega$ reaches the value of
$\half S/\Omega$ for different values of $S/\Omega$.
}
\label{FMRI_mountains}
\end{figure}

In \Fig{FMRI_bump} we show the evolution of $\mbox{Re}\lambda$ as
a function of time, except that time is translated into a corresponding
variation of $k_x(t)$; see \Eq{kxt}.
In the second panel we show the corresponding variation of
$\bra{\qq|\qq}^{1/2}$, where we see an increase over about 8 orders
of magnitude during the time interval in which $\mbox{Re}\lambda$ is
systematically positive.
Note also that $\max(\mbox{Re}\lambda)\approx{3\over4}\Omega$
(upper panel of \Fig{FMRI_bump}),
which is indeed the maximum growth rate of the axisymmetric MRI.
This supports our interpretation that the Rayleigh quotient
provides a convenient and quantitative means of estimating the
growth rate of the instability.

In \Fig{FMRI_mountains} we show a two-dimensional parameter survey
in the $(k,S)$ plane of the gain factor and of
$\max(\mbox{Re}\lambda)$.
Note that both the gain factor and $\max(\mbox{Re}\lambda)$ increase
toward more negative values of $S$.
Remarkable is the fact that on the Rayleigh line, $S/\Omega=-2$,
both quantities vary smoothly and do not show any special behavior.

\subsection{Direct numerical verification}

\begin{figure}[t!]
\includegraphics[width=\columnwidth]{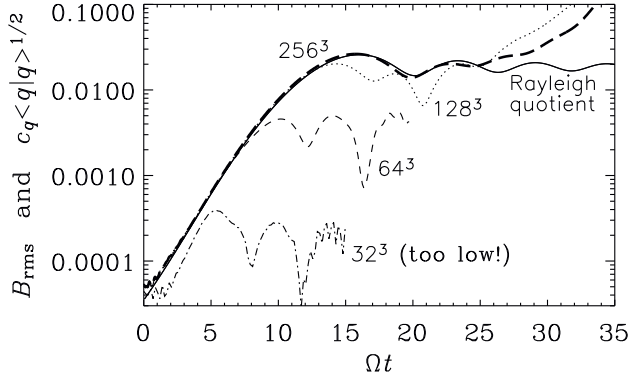}
\caption{Transient amplification of the magnetic field by the
nonaxisymmetric MRI with an azimuthal field.
The solid line shows the result from the Rayleigh quotient method
while the broken lines give the result from direct three-dimensional
simulations with zero viscosity and zero resistivity.
The square root of the Rayleigh quotient has been scaled by a factor
$c_q=3.5\times10^{-5}$ to make it overlap with $B_{\rm rms}$ curve.
A resolution of only $32^3$ meshpoints is completely insufficient to
resolve even the beginning of the instability.
At least $256^3$ meshpoints are required to resolve the maximum
(thick dashed line).
After $\Omega t>17$ even the simulation with $256^3$ meshpoints
becomes under-resolved.}
\label{pcomp_col}
\end{figure}

It is instructive to compare the present modal analysis with
a direct three-dimensional simulation of the fully nonlinear
equations in real space.
This is done in \Fig{pcomp_col} where we compare the increase of
$\bra{\qq|\qq}^{1/2}$ with the resulting evolution of the
root-mean-square magnetic field from a direct simulation
of the shearing sheet equations. Here we also adopt an isothermal
gas with constant sound speed.

\begin{figure*}[t]
\includegraphics[width=\textwidth]{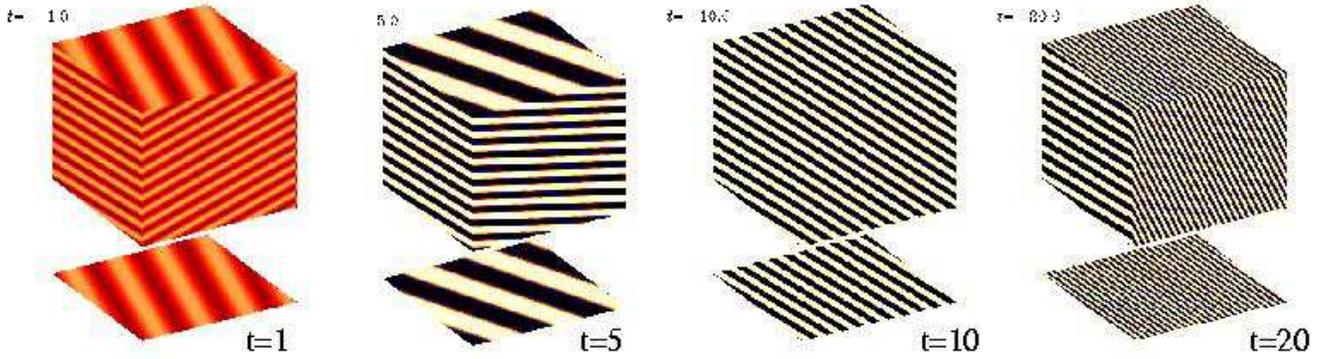}
\caption{ 
Images of the vertical component of the magnetic field, $B_z$, for
the nonaxisymmetric MRI with an azimuthal field, using
different values of $t$, where the field at $t=0$ corresponds to that
given by \Eq{initcondUB}.
}\label{img}\end{figure*}

The initial condition for the 3-dimensional direct simulation is obtained
by evolving linearized shearing sheet equations for $k_y=1$ and $k_z=10$
to the point where $k_x(t_0)=-5$. [The size of the domain is $(2\pi)^3$.]
For definitiveness, we reproduce here the numerical values in
\Eq{expansion}:
\EQ
\hat{\uu}\!=\!\pmatrix{
-0.311-0.037\ii\cr
-0.461-0.054\ii\cr
-0.088-0.010\ii}\!,
\quad
\hat{\bb}\!=\!\pmatrix{
+0.068-0.581\ii\cr
-0.039+0.334\ii\cr
+0.035-0.295\ii}\!,
\label{initcondUB}
\EN
and the logarithmic density is given by $\hat{\Lambda}=0.042-0.3647\ii$.
The amplitude is chosen to be $A=10^{-4}$.
\FFig{img} shows images of $B_z$ on the periphery of the simulation domain
at different times\footnote{Animations of the flow can be found on
\url{http://www.nordita.dk/~brandenb/movies/disc}}.
The simulations have been carried out using the {\sc Pencil Code}\footnote{
\url{http://www.nordita.dk/software/pencil-code}} which is a high-order
finite-difference code (sixth order in space and third
order in time) for solving the compressible hydromagnetic equations.

The way how this transient amplification can lead to sustained
growth is through mode coupling, which is not considered in the present
analysis.
Relevant mode couplings could come about either through nonuniformities
in the cross-stream or $x$ direction and through boundary conditions, or through
nonlinearities. 
In the shearing sheet approximation only the latter is a viable
possibility, and this is probably the mechanism through which the early
shearing sheet simulations produced sustained turbulence (e.g.\ Hawley et 
al.\ 1995, Matsumoto \& Tajima 1995, Brandenburg et al.\ 1995).

We conclude this section by emphasizing the one-to-one correspondence
between numerical solutions of the shearing sheet equations and the
Rayleigh quotient method.
The nonaxisymmetric case is similar to the axisymmetric one in that
the maximum growth rate is the same.
This is achieved for $k_x=0$, which, in the nonaxisymmetric case,
can only be the case at one instance in time.

\section{Shear flow with stratification}

We now turn to the case with vertical stratification.
In order to allow for gravity waves, we need to abandon the
isothermal equation of state and use instead a perfect gas law.
In terms of specific entropy, $s$, we can formulate this as
\EQ
s=c_v\ln p-c_p\ln\rho+s_0,
\label{eosSRI}
\EN
where $c_p$ and $c_v$ are the specific heats at constant
pressure and constant volume, respectively.
The specific entropy is only defined up to an additive constant, $s_0$.
In the absence of heating and cooling, the specific entropy
remains constant on streamlines, i.e.\
\EQ
{\DD s\over\DD t}=0.
\label{dsdt}
\EN
Using \Eqs{eosSRI}{dsdt} we can rewrite the continuity equation
\eq{drhodt} in the form
\EQ
{\DD p\over\DD t}=-\gamma p\nab\cdot\uu,
\label{dpdtSRI}
\EN
where $\gamma=c_p/c_v$ is the ratio of specific heats
and $\nab\cdot\meanUU=0$ has been used.
We solve \Eqs{dsdt}{dpdtSRI} together with the momentum
equation in the form
\EQ
\rho{\DD\uu\over\DD t}
=\rho\ff(\uu)-\nab p+\rho\grav,
\label{dudtSRI}
\EN
where $\ff(\uu)$ has been defined in \Eq{ffuuDefn}.
These equations have an isothermal equilibrium solution
(denoted by an overbar),
\EQ
\meanuu=\vec{0},\quad
\overline{s}(z)={c_p-c_v\over H},\quad
\overline{p}(z)=p_0 e^{-z/H},
\EN
where
\EQ
H=c_{\rm s}^2/(\gamma g)
\EN
is the pressure scale height and $c_{\rm s}$ is the sound speed.
Both $H$ and $c_{\rm s}$ are constants, and so is $p_0$ which
gives the pressure at $z=0$.
The equilibrium density is given by
$\overline{\rho}=\gamma\overline{p}/c_{\rm s}^2$.

Linearizing \Eqss{eosSRI}{dudtSRI} about the
$z$ dependent equilibrium solution yields
\EQ
{s'\over c_p}=
{p'\over\gamma\overline{p}}
-{\rho'\over\overline{\rho}},
\label{eosSRIlin}
\EN
\EQ
{\partial s'\over\partial t}+u_z'{\dd\overline{s}\over\dd z}=0,
\EN
\EQ
{\partial p'\over\partial t}+u_z'{\dd\overline{p}\over\dd z}
=-\gamma\overline{p}\nab\cdot\uu',
\EN
\EQ
\overline\rho(z){\partial\uu'\over\partial t}=
\overline\rho(z)\ff(\uu')-\nab p'+\rho'\grav,
\EN
where primes denote deviations from the equilibrium.
These four equations have non-constant coefficients.
Using rescaled variables (denoted by a tilde),
\EQ
s'=\sqrt{\gamma-1}\,c_p\tilde{s} e^{z/2H},
\EN
\EQ
p'=\gamma p_0\tilde{p} e^{-z/2H},
\EN
\EQ
\uu'=c_{\rm s}\tilde{\uu} e^{z/2H},
\EN
and eliminating $\rho'$ using \Eq{eosSRIlin},
we can rewrite these equations in the form
\EQ
{\partial\tilde{s}\over\partial t}
=-\sqrt{\gamma-1}\,{c_{\rm s}\over\gamma H}\tilde{u}_z
\label{EntropyWithGamma}
\EN
\EQ
{\partial\tilde{p}\over\partial t}
={c_{\rm s}\over\gamma H}\tilde{u}_z
-c_{\rm s}\left({\tilde{u}_z\over2H}+\nab\cdot\tilde{\uu}\right)
\label{ContinuityWithGamma}
\EN
\EQ
{\partial\tilde{\uu}\over\partial t}
=\ff(\tilde{\uu})
+c_{\rm s}\left({\zz\over2H}-\nab\right)\tilde{p}
+\left(\tilde{p}-\sqrt{\gamma-1}\,\tilde{s}\right)
{\grav\over c_{\rm s}},
\label{MomentumWithGamma}
\EN
It is now convenient to introduce nondimensional time and
space coordinates.
A natural length scale to choose might be $H$, or better
$\gamma H=c_{\rm s}^2/g$.
The problem with this is that the unstratified limit $g\to0$
is then ill-posed, because $H\to\infty$.
A more flexible alternative is therefore to choose a ``system''
scale $L$, or better $\gamma L$, as our length scale.
(The length scale $L$ is not to be confused with the matrix $\LLLL$.)
Thus, we define nondimensional inverse time and length scales via
\EQ
{\partial\over\partial t}={c_{\rm s}\over\gamma L}
{\partial\over\partial\tilde{t}}\quad\mbox{and}\quad
\nab={1\over\gamma L}\tilde{\nab}.
\EN
The final set of equations is then
\EQ
{\partial\tilde{s}\over\partial\tilde{t}}
=-{L\over H}\sqrt{\gamma-1}\,\tilde{u}_z,
\label{EntropyWithGammaDimless}
\EN
\EQ
{\partial\tilde{p}\over\partial\tilde{t}}
={L\over H}\left(1-{\gamma\over2}\right)\tilde{u}_z
-\tilde{\nab}\cdot\tilde{\uu},
\label{ContinuityWithGammaDimless}
\EN
\EQ
{\partial\tilde{\uu}\over\partial\tilde{t}}
=\tilde{\ff}(\tilde{\uu})
-{L\over H}\left(1-{\gamma\over2}\right)\zz\tilde{p}
-\tilde{\nab}\tilde{p}
+{L\over H}\sqrt{\gamma-1}\,\zz\tilde{s},
\label{MomentumWithGammaDimless}
\EN
where we have defined
$\tilde{\ff}(\tilde{\uu})=(2\tilde{\Omega}u_y,-2\tilde{\Omega}^Su_x,0)^T$
with $2\tilde{\Omega}^S=2\tilde{\Omega}+\tilde{S}$.
Here, $\tilde{\Omega}=\gamma L\Omega/c_{\rm s}$ and
$\tilde{S}=\gamma L S/c_{\rm s}$ are nondimensional
angular and shear velocities.
We recall that $\grav=(0,0,-g)$.
As a nondimensional measure for the degree of stratification we use
in the following the symbol
\EQ
\tilde{g}={\gamma L g\over c_{\rm s}^2}\equiv{L\over H}.
\EN
The case $\tilde{g}=0$ corresponds obviously to the completely
unstratified case.

Next, we write these equations in matrix form \eq{dqdt}, make the ansatz
\eq{expansion}, and have a set of ordinary differential equations \eq{ode}
with the matrix $\hat{\LLLL}$ in the form
\EQ
\hat{\LLLL}=\pmatrix{
0&2\tilde{\Omega}&0&0&-\ii k_x(t)\cr
-2\tilde{\Omega}^S&0&0&0&-\ii k_y\cr
0&0&0&\tilde{N}&-\tilde{M}-\ii k_z\cr
0&0&-\tilde{N}&0&0\cr
-\ii k_x(t)&-\ii k_y&\tilde{M}-\ii k_z&0&0},
\label{LLLLhat}
\EN
operating on the state vector
$\hat{\qq}=(\hat{u}_x,\hat{u}_y,\hat{u}_z,\hat{s},\hat{p})^T$,
which is defined according to \Eq{expansion},
\EQ
\tilde{N}={L\over H}\sqrt{\gamma-1}
\EN
is the nondimensional Brunt-V\"ais\"al\"a frequency, and
\EQ
\tilde{M}={L\over H}\left(1-{\gamma\over 2}\right)
\EN
is another nondimensional number characterizing the degree of stratification.
Note that in the absence of shear, $S=0$, the matrix $\hat{\LLLL}$
is skew-hermitian, i.e.\ $\hat{\sf L}_{ij}=-\hat{\sf L}_{ji}^*$,
and all eigenvalues are purely imaginary or zero.
A similar nondimensionalization was used by Brandenburg (1988) for the case
$\Omega=0$, but space and time coordinates were scaled such that the
coefficient $\tilde{M}$ became unity.
As discussed above, this gives an unnecessary restriction in that it
prevents us from making a continuous transition to the unstratified
case $L/H=0$.
Furthermore, in the special case $\gamma=2$, we have $\tilde{M}=0$,
$\tilde{N}=L/H\neq0$, which corresponds to the Boussinesq case if we
also make the assumption of incompressibility.
Technically, the latter is achieved by multiplying the left hand side
of \Eq{ode} by the matrix ${\rm diag}(1,1,1,1,0)$.
This means that $\tilde{p}$ can be expressed in terms of $\tilde{\uu}$,
and we are left with only 4 explicitly time-dependent equations.

In the axisymmetric case, $k_y=0$, or in the absence of shear, $S=0$,
the eigenvalues $\lambda$ satisfy the dispersion relation
$\lambda D(\lambda,\kk)=0$ with
\EQ
D=\lambda^4+\lambda^2(k^2+\kappa^2+\tilde{K}^2)
+k_x^2\tilde{N}^2+\kappa^2(k_z^2+\tilde{K}^2),
\label{disper_atmos}
\EN
where $k^2=k_x^2+k_z^2$, and $\kappa$ is
the nondimensional epicyclic frequency with
$\kappa^2=4\tilde\Omega\tilde\Omega^S$, and $\tilde{K}^2=\tilde{N}^2+\tilde{M}^2$
has been introduced for abbreviation.
Written in dimensional form, this dispersion relation is identical
to the usual one for atmospheric waves in an isothermally stratified
atmosphere (Stein \& Leibacher 1974, Ryu \& Goodman 1992). There are
five solutions, one zero-frequency mode, $\lambda=0$, and four others
with $\mbox{Im}\,\lambda=\pm\omega_{\rm p}(\kk)$ and
$\mbox{Im}\,\lambda=\pm\omega_{\rm g}(\kk)$,
corresponding to p- and g-modes, respectively.

Only in the
{\it axisymmetric} case, $k_y=0$, $\hat{\LLLL}$ becomes independent of $t$
($k_x=k_{x0}={\rm const}$) and so \Eq{disper_atmos} can be used even when
$S\neq0$. From \Eq{disper_atmos} it is clear that now instability requires
\EQ
\kappa^2<-k_x^2\tilde{N}^2/(k_z^2+\tilde{K}^2),
\EN
which is also known as the Solberg-H{\o}iland criterion
(e.g.\ R\"udiger \& Shalybkov 2002, Narayan et al.\ 2002).
Negative values of $\kappa^2=4\tilde\Omega\tilde\Omega^S$ are possible
when $-S>2\Omega$, corresponding to Rayleigh's criterion.
Instability occurs closest to the Rayleigh line when
$k_x^2$ is small values and $k_z^2$ large.

\begin{figure}[t!]
\centering\includegraphics[width=\columnwidth]{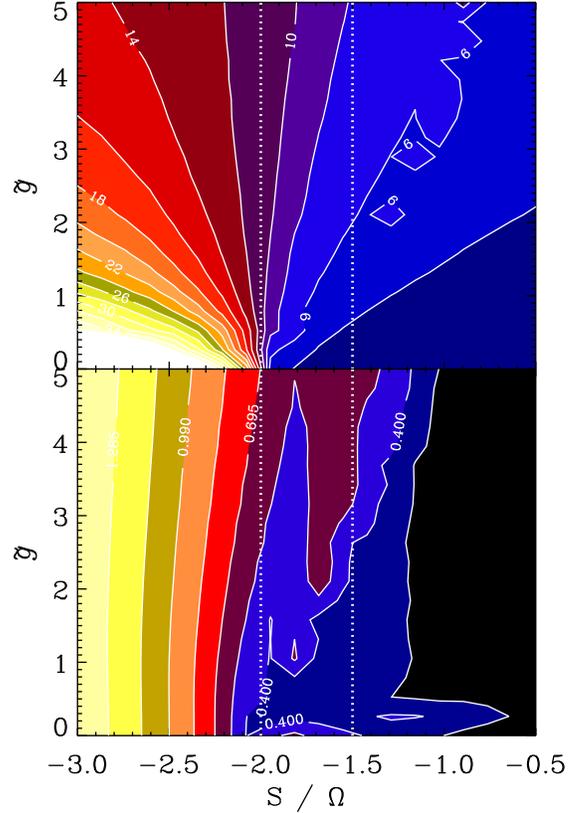}
\caption{Nonaxisymmetric stratified case:
dependence of the gain factor in linear representation
(upper panel) and $\max_t(\lambda)/\Omega$ (lower panel)
on $S/\Omega$ and $\tilde{g}$, for $\gamma L k_z=1$.}
\label{SRI_nonaxi}
\end{figure}

In \Fig{SRI_nonaxi} we present the results of a two-dimensional
parameter survey varying both the degree of stratification,
$\tilde{g}$ as well as the strength of the shear, $S$.
The Rayleigh stability line, $S/\Omega=-2$, is particularly
evident in the plot of the gain factor (upper panel): for
zero stratification the gain factor increases sharply as one
crosses the Rayleigh line.
For stronger stratification, the increase in the gain factor
diminishes, suggesting that stratification has a stabilizing
influence on the Rayleigh-Taylor instability.

\begin{figure}[t!]
\centering\includegraphics[width=\columnwidth]{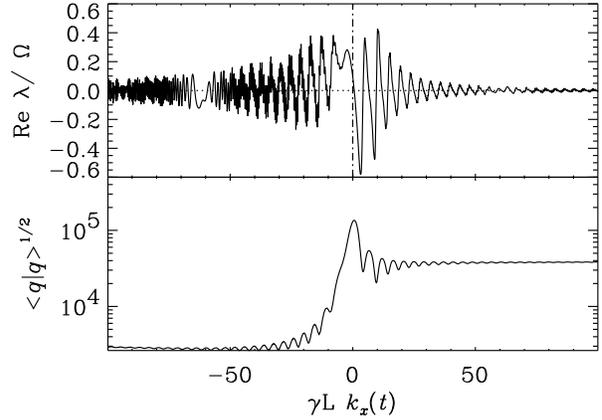}
\caption{Nonaxisymmetric stratified case:
dependence of the gain factor and $\max_t(\lambda)/\Omega$
on $\gamma L k_x(t)$, for $\gamma L k_z=1$.}
\label{SRI_growth}
\end{figure}

For $S/\Omega>-2$, the flow is known to be axisymmetrically stable.
Also in the nonaxisymmetric case the gain factor is negligible.
Nevertheless, in the $(\tilde{g},S)$ parameter plane there are
regions with an appreciable maximum value of $\lambda$ (lower
panel of \Fig{SRI_nonaxi}).
However, as is seen more clearly in the corresponding time trace
(\Fig{SRI_growth}), a positive maximum of $\lambda$ is more or
less compensated by a corresponding amount of negative contributions.
This explains the absence of a corresponding pattern of
$\max_t(\lambda)$ in the $(\tilde{g},S)$ plane.
This suggests also that, in contrast to simulations of the MRI,
the transients would be insignificant to produce sustained growth.

\section{Conclusions}

The shearing sheet approximation is a powerful tool to analyze the
behavior of an accretion disc locally.
While this approximation eliminates a wide class of global instabilities, it has
the advantage of isolating local instabilities that are often believed
to be responsible for driving turbulence.
Our work has demonstrated that for sufficiently weak perturbations
there is a direct correspondence between the solutions of shearing sheet
equations as obtained from a numerical code and the Rayleigh quotient
obtained by solving the modal equations with a time-dependent radial
wavenumber $k_x(t)$.

Within the framework of the shearing sheet approximation we
have demonstrated the stability of a stratified isothermal atmosphere
with horizontal keplerian shear to fully compressible perturbations.
The behavior of the solution is well characterized by the Rayleigh
quotient, $\lambda(t)$, which is
obtained for a range of different initial conditions.
The strato-rotational instability (SRI), which has been seen in experiments
and calculations with radial boundaries (Molemaker et al.\ 2001,
Shalybkov \& R\"udiger 2005), can therefore not be a local instability,
as was already discussed by Umurhan (2005) and Lesur \& Longaretti (2005).

The present work has also highlighted the absence of any correspondence between
the growth rates obtained from the {\it pseudo}-dispersion relation and
the actual evolution of $\mbox{Re}\,\lambda(t)$. Of course, there will be
transient growth from almost any initial condition and for any value of
$S/\Omega$ -- including $S/\Omega>0$, which would be stable even in the
presence of magnetic fields (Balbus \& Hawley 1991).
However, under certain conditions this can lead to what is known as the
bypass transition (Chagelishvili et al.\ 2003, Afshordi et al.\ 2005),
i.e.\ the transient growth that can lead to a new state that is itself unstable.
Nevertheless, the presence of
stratification ($\tilde{g}>0$) is here seen to have
a stabilizing effect, making the connection with the SRI implausible.

We emphasize that in the {\it non-Boussinesq} stratified
case it is important to remove $z$-dependent coefficients before making
the $\exp(\ii k_zz)$ ansatz. Otherwise, persistent real parts
of $\lambda(t)$ will arise that reflect merely the fact that velocity
increases as a wave packet travels into less dense regions.

Finally, it should be pointed out that, if keplerian shear flows were
locally unstable in stratified media, one might have seen this in the
fully nonlinear shearing sheet simulations of Brandenburg \ea (1995),
where the Lorentz force was removed and the dynamo-driven turbulence
was found to decay rapidly; see their Fig.~4. A similar test
was also done by Hawley \ea (1995), but only for the unstratified case
which is not relevant here.
We note that real discs have vertical shear which does allow for a
hydrodynamic instability, but it is far less powerful than
the Balbus-Hawley instability (e.g.\ Urpin \& Brandenburg 1998,
Arlt \& Urpin 2004).
Other proposals for hydrodynamic instability include the baroclinic
instability (Klahr \& Bodenheimer 2003) that has been discussed in
the context of protostellar discs.

\begin{acknowledgements}
We thank B\'ereng\`ere Dubrulle for numerous email conversations
on the subject of this paper.
This work has been supported by the European Commission under
the Marie-Curie grant HPMF-CT-1999-00411.
The Danish Center for Scientific Computing is acknowledged for
granting time on the Horseshoe cluster.
\end{acknowledgements}


\begin{thebibliography}{99}

\bibitem[]{}
Arlt, R., \& Urpin, V.\yana{2004}{426}{755}

\bibitem[]{}
Afshordi, N., Mukhopadhyay, B., \& Narayan, R.\yapj{2005}{629}{373}

\bibitem[]{}
Balbus, S. A., \& Hawley, J. F.\yapj{1991}{376}{214}

\bibitem[]{}
Balbus, S. A., \& Hawley, J. F.\yapj{1992a}{400}{610}

\bibitem[]{}
Balbus, S. A., \& Hawley, J. F.\yapj{1992b}{392}{662}

\bibitem[]{}
Balbus, S. A., \& Hawley, J. F.\yjour{1998}{Rev. Mod. Phys.}{70}{1}

\bibitem[]{}
Brandenburg, A.\yana{1988}{203}{154}

\bibitem[]{}
Brandenburg, A., \& R\"udiger, G.\sana{2006}, arXiv: astro-ph/0512409.

\bibitem[]{}
Brandenburg, A., Nordlund, \AA., Stein, R. F., \& Torkelsson, U.
\yapj{1995}{446}{741}

\bibitem[]{}
Chagelishvili, G. D., Zahn, J.-P., Tevzadze, A. G., \&
Lominadze, J. G.\yana{2003}{402}{401}

\bibitem[]{}
Dubrulle, B., Mari\'e, L., Normand, Ch., Richard, D., Hersant, F., \&
Zahn, J.-P.\yana{2005}{429}{1}

\bibitem[]{}
Frank, J., King, A. R., \& Raine, D. J.\ybook{1992}{Accretion power
in astrophysics}{Cambridge: Cambridge Univ. Press}

\bibitem[]{}
Goldreich, P., \& Lynden-Bell, D.\ymn{1965}{130}{125}

\bibitem[]{}
Hawley, J. F., Gammie, C. F., \& Balbus, S. A.\yapj{1995}{440}{742}

\bibitem[]{}
Kim, W.-T., \& Ostriker, E. C.\yapj{2000}{540}{372}

\bibitem[]{}
Klahr, H., \& Bodenheimer, P.\yapj{2003}{582}{869}

\bibitem[]{}
Lesur, G., \& Longaretti, P.-Y.\yana{2005}{444}{25}

\bibitem[]{}
Matsumoto, R., \& Tajima, T.\yapj{1995}{445}{767}

\bibitem[]{}
Molemaker, M.J., McWilliams, J.C., \& Yavneh, I.\yprl{2001}{86}{5270}

\bibitem[]{}
Narayan, R., Quataert, E., Igumenshchev, I. V.,
\& Abramowicz, M. A.\yapj{2002}{577}{295}

\bibitem[]{}
Ogilvie, G. I., \& Pringle, J. E.\ymn{1996}{279}{152}

\bibitem[]{}
R\"udiger, G., \& Shalybkov, D.\ypre{2002}{66}{016307}

\bibitem[]{}
Ryu, D., \& Goodman, J.\yapj{1992}{388}{438}

\bibitem[]{}
Shakura, N. I., \& Sunyaev, R. A.\yana{1973}{24}{337}

\bibitem[]{}
Shalybkov, D., \& R\"udiger, G.\yana{2005}{438}{411}

\bibitem[]{}
Stein, R. F., \& Leibacher, J.\yanar{1974}{12}{407}

\bibitem[]{}
Umurhan, O. M.\ymn{2006}{365}{85}

\bibitem[]{}
Urpin, V., \& Brandenburg, A.\ymn{1998}{294}{399}

\end{thebibliography}
\end{document}